\documentclass[fleqn,usenatbib]{mnras}
\usepackage{newtxtext,newtxmath}
\usepackage{xcolor}
\usepackage[T1]{fontenc}
\usepackage{ae,aecompl}
\usepackage{graphicx}
\usepackage{amsmath}	
\usepackage{amssymb}	

\usepackage{bm}

\title[Gravitational lensing of gravitational waves]{Gravitational
  lensing of gravitational waves: wave nature and prospects for
  detection}  

\author[Meena and Bagla]{
Ashish Kumar Meena,$^{1}$\thanks{E-mail: ashishmeena@iisermohali.ac.in}
Jasjeet Singh Bagla,$^{1}$\thanks{E-mail: jasjeet@iisermohali.ac.in}
\\
$^{1}$Indian Institute of Science Education and Research Mohali,
Knowledge City, Sector 81, Sahibzada Ajit Singh Nagar, Punjab 140306,
India}

\pubyear{2019}

\begin{document}
\label{firstpage}
\pagerange{\pageref{firstpage}--\pageref{lastpage}}
\maketitle

\begin{abstract}
  We discuss the gravitational lensing of gravitational wave signals
  from coalescing binaries.
  We delineate the regime where wave effects are significant from the
  regime where geometric limit can be used.
  Further, we focus on the effect of micro-lensing and the combined
  effect of strong lensing and micro-lensing.
  We find that micro-lensing combined with strong lensing can introduce
  time varying phase shift in the signal and hence can lead to
  detectable differences in the signal observed for different images
  produced by strong lensing.
  This, coupled with the coarse localization of signal source in the
  sky for gravitational wave detections, can make it difficult to
  identify the common origin of signal corresponding to different
  images and use observables like time delay.
  In case we can reliably identify corresponding images, micro-lensing
  of individual images can be used as a tool to constrain properties
  of micro-lenses.
  Sources of gravitational waves can undergo microlensing due to
  lenses in the disk/halo of the Galaxy, or due to lenses in an
  intervening galaxy even in absence of strong lensing.
  In general the probability for this is small with one exception: 
  Extragalactic sources of gravitational waves that lie in the galactic
  plane are highly likely to be micro-lensed.
  Wave effects are extremely important for such cases.
  In case of detections of such sources with low SNR, 
  the uncertainty
  of occurrence of microlensing or otherwise introduces an additional 
  uncertainty in the parameters of the source.
\end{abstract}


%%%%%%%%%%%%%%%%% BODY OF PAPER %%%%%%%%%%%%%%%%%%

\section{Introduction}
\label{sec:Introduction}

The recent detection of gravitational wave (GW) signal
\citep{Abbott_2018} from coalescing binaries opens up a new window to
observe the Universe \citep{rosswog_2015,wei_2017}.  
The upcoming runs of LIGO and Virgo with increased sensitivity and new
detector facilities \citep{KAGRA_2018} will increase the number of
observed GW signals significantly.  
As a result, the possibility of detecting a gravitationally lensed GW
will also increase.  

Propagation of gravitational waves is influenced by gravitational
fields and results in a deflection in a manner similar to
electromagnetic waves \citep{Lawrence_1971_2,Ohanian_1974}. 
In the case of electromagnetic radiation, geometric optics is
sufficient to study the effects of gravitational lensing as the
sources have a finite size and the wavelength of the radiation is much
smaller than all other scales of interest
\citep{1986PhRvD..34.1708D,Schneider_1992,2006JCAP...01..023M}. 
Sources of gravitational waves are fairly compact and the wavelength
of radiation that can be detected by existing and future detectors is
larger than the region of emission.
The wavelength of gravitational waves is comparable with the
Schwarzschild radius of many astronomical objects.
Thus in the case of gravitational waves, the geometric optics is not
always valid
\citep{Bontz_1981,Deguchi_1986,Nakamura_1998,Baraldo_1999,
  2018PhRvD..98j3022C}.    
In LIGO frequency band ($10Hz-10kHz$) for galaxy mass lenses geometric
optics is sufficient \citep{Varvella_2004} as the wavelengths are much
smaller than the size of a galaxy. 
Under geometric optics approximation, for galaxy mass lenses, the GW
signal gets multiply imaged, and each signal is amplified by a
constant factor. 
Although extra amplification can help us in observing events that are
beyond LIGO range, the amplification factor and luminosity distance to
the source have a degeneracy that can introduce errors in the analysis
\citep{Broadhurst_2018,Broadhurst_2019}.
If we can identify the multiple {\sl images} in a lensed system then
the time delay between these different signals can constrain the
cosmological parameters \citep{Sereno_2011}.  
The presence of small compact objects (micro-lenses) in the lens can
further affect the signal.
In this case the wave nature can become important and hence the effect
of lensing is a combination of a wavelength dependent amplification
factor as well as the phase of the signal~\citep{Diego_2019}.

In this work we discuss the effects of micro-lensing in strongly lensed
GW signal.
The effects of micro-lensing are non-negligible in the LIGO frequency
band~\citep{2018PhRvD..98j3022C,Diego_2019}. 
As we shall see, these effects introduce frequency dependence in
amplification as well as the phase of the lensed GW signal.  

In section \S{\ref{sec:lensing}}, we review the basics of wave optics
in gravitational lensing.  
Results are given in \S{\ref{sec:results}}. 
Summary and conclusions are given in \S{\ref{sec:conclusions}}.
We also discuss possibilities for future work in this section.

%%%%%%%%%%%%%%%%%%%%%%%%%%%%%%%%%%%%%%
\begin{figure*}
\includegraphics[width=12cm, height= 8cm]{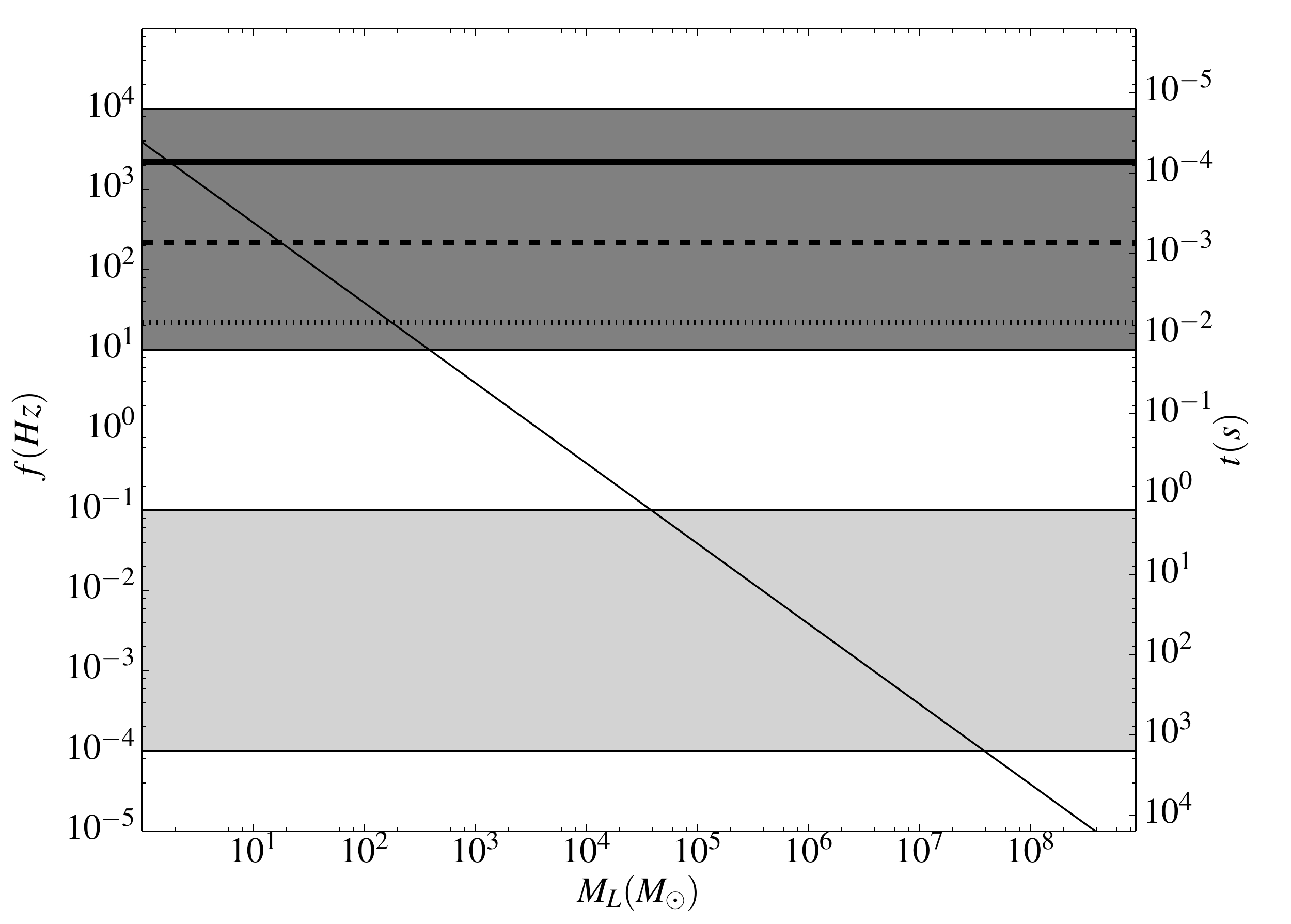}
\caption{Lens mass versus gravitational wave frequency: The diagonal
  thin line from top left to bottom right represents the critical
  frequency below which wave effects are essential to calculate the
  lensing effects for a given point mass.   
  The light and dark shaded regions represent the relevant frequencies
  for LISA and LIGO, respectively.
  The horizontal lines (dotted, dashed, thick solid line) represents
  the GW frequency at the innermost stable circular orbit (ISCO)
  emitted by the source binaries of masses, $100 M_\odot+100M_\odot$,
  $10M_\odot+10M_\odot$, $1M_\odot+1M_\odot$, respectively.}  
\label{fig:cutoff_frequency}
\end{figure*}
%%%%%%%%%%%%%%%%%%%%%%%%%%%%%%%%

\section{Basics of Gravitational Lensing}
\label{sec:lensing}

In this section, we review some aspects of wave effects in
gravitational lensing, which is relevant for our 
analysis~\citep{Schneider_1992,Nakamura_1999,Takahashi_2003}.  
To describe the effect of the gravitational lens, one can consider a
perturbed FRW as background metric, given by 
\begin{equation}
ds^2=-(1+2U)\:
dt^2+a^2(1-2U)\:d\mathbf{r}^2=g_{\mu\nu}^{\left(B\right)}dx^\mu
dx^\nu, 
\end{equation}
where $U$ is the gravitational potential of the lens.
The gravitational waves in this background are described as a tensor
perturbation, $h_{\mu\nu}$.
The linear perturbation can be written as $\phi e_{\mu\nu}$; where
$\phi$ represents the amplitude, and $e_{\mu\nu}$ is the polarization
tensor of the gravitational wave.
During propagation, the change in polarization tensor due to the
presence of the lens is negligible \citep{Takahashi_2003,Misner_1973}. 
As a result, one can assume the polarization vector to be a constant.
The propagation equation (to the leading order) for the scalar
amplitude $\phi$, in frequency domain is,
\begin{equation}
\left(\nabla^2 +\widetilde{\omega}^2\right)\widetilde{\phi}=4
\widetilde{\omega}^2U\widetilde{\phi}, 
\label{eq:scalar_amp}
\end{equation}
where $\widetilde{\omega}=2\pi f$, and $f$ is the frequency of the
gravitational wave.
One can solve the equation \eqref{eq:scalar_amp} using Kirchhoff
diffraction integral~\citep{Baraldo_1999, Takahashi_2003} for waves
coming from a source at a distance $D_S$ to the observer.
Following~\cite{Takahashi_2003}, the amplification factor,
$F\left(f\right)$ is defined as the ratio of lensed and unlensed
$\left(U=0\right)$ gravitational wave amplitudes $\widetilde{\phi}$.
This implies that in no-lens limit $\left(U=0\right)$, the
amplification factor is unity $\left(|F|=1\right)$.  

In order to calculate the amplification factor, we need to solve the
equation \eqref{eq:scalar_amp} for a given lens system.
Under thin lens approximation, for a gravitational wave source at a
distance $D_S$ and a lens at $D_L$, the amplification factor at the
observer is given by~\cite{Takahashi_2003}, 
\begin{equation}
F\left(f,\mathbf{y}\right)=\frac{D_S \xi_0^2 \left(1+z_L\right)}{c D_L
  D_{LS}}\frac{f}{\textit{i}} \int d^2\mathbf{x}\:
\exp\left[2\pi\textit{i} f
  t_d\left(\mathbf{x},\mathbf{y}\right)\right], 
\label{eq:amplification_factor}
\end{equation}
where $\mathbf{x}=\bm{\xi}/\xi_0$, $\mathbf{y}=\bm{\eta}D_L/\xi_0D_S$
is the dimensionless source position in the source plane, $\xi_0$ is
an arbitrary length scale, $z_L$ is the lens redshift and $t_d$ is the
arrival time, given by 
\begin{equation}
t_d\left(\mathbf{x},\mathbf{y}\right) = \frac{D_S\xi_0^2(1+z_L)}{c D_{LS}
  D_S}\left[\frac{1}{2}|\mathbf{x}-\mathbf{y}|^2 -
  \psi\left(\mathbf{x}\right)+\phi_m\left(\mathbf{y}\right) \right]. 
\end{equation}
The constant $\phi_m\left(\mathbf{y}\right)$ does not depend on the
lens properties and one can choose a form to simplify calculations. 
Here we choose $\phi_m\left(\mathbf{y}\right)$ such that the value of
time delay for minima image (image for which time delay is a global
minimum) is zero: time delays for other points are measured
with respect to this image. 
The phase of the amplification factor can be defined as,
$\triangle\phi = -\textit{i}~\ln\left[F/|F|\right]$. 

In the limit of geometric optics approximation ($f \gg t_{d}^{-1}$), the
integral in equation  \eqref{eq:amplification_factor} is a highly
oscillatory function, and the amplification factor gets a significant
contribution only from stationary points of the integrand. 
The form of amplification factor in geometric optics limit is 
\begin{equation}
F\left(f,\mathbf{y}\right)=\sum_i\sqrt{|\mu_j|}\exp\left(2\pi
  \textit{i} f t_{d,j}-\textit{i}\pi n_j\right), 
\end{equation}
where $t_{d,j}$ is the value of time delay for $j$-th image, $\mu_j$
is the amplification of the $j$-th image and $n_j$ is the Morse index
with values 0, 1/2, 1 for images corresponding to minima, saddle and
maxima of the time delay function,
respectively~\citep{Takahashi_2003,Blandford_1986}.  
One can see that the lensed gravitational waveform corresponding to
maxima (saddle) of the time delay has a phase difference of
$e^{-\textit{i}\pi}$($e^{\frac{-\textit{i}\pi}{2}}$) compared to the
waveform corresponding to the minima of time delay~\citep{Dai_2017}. 

%%%%%%%%%%%%%%%%%%%%%%%%%%%%%%%%%%%%%%%%%%%%%%%%%%%%%%%
\begin{figure*}
 \includegraphics[width=8cm, height= 6cm]{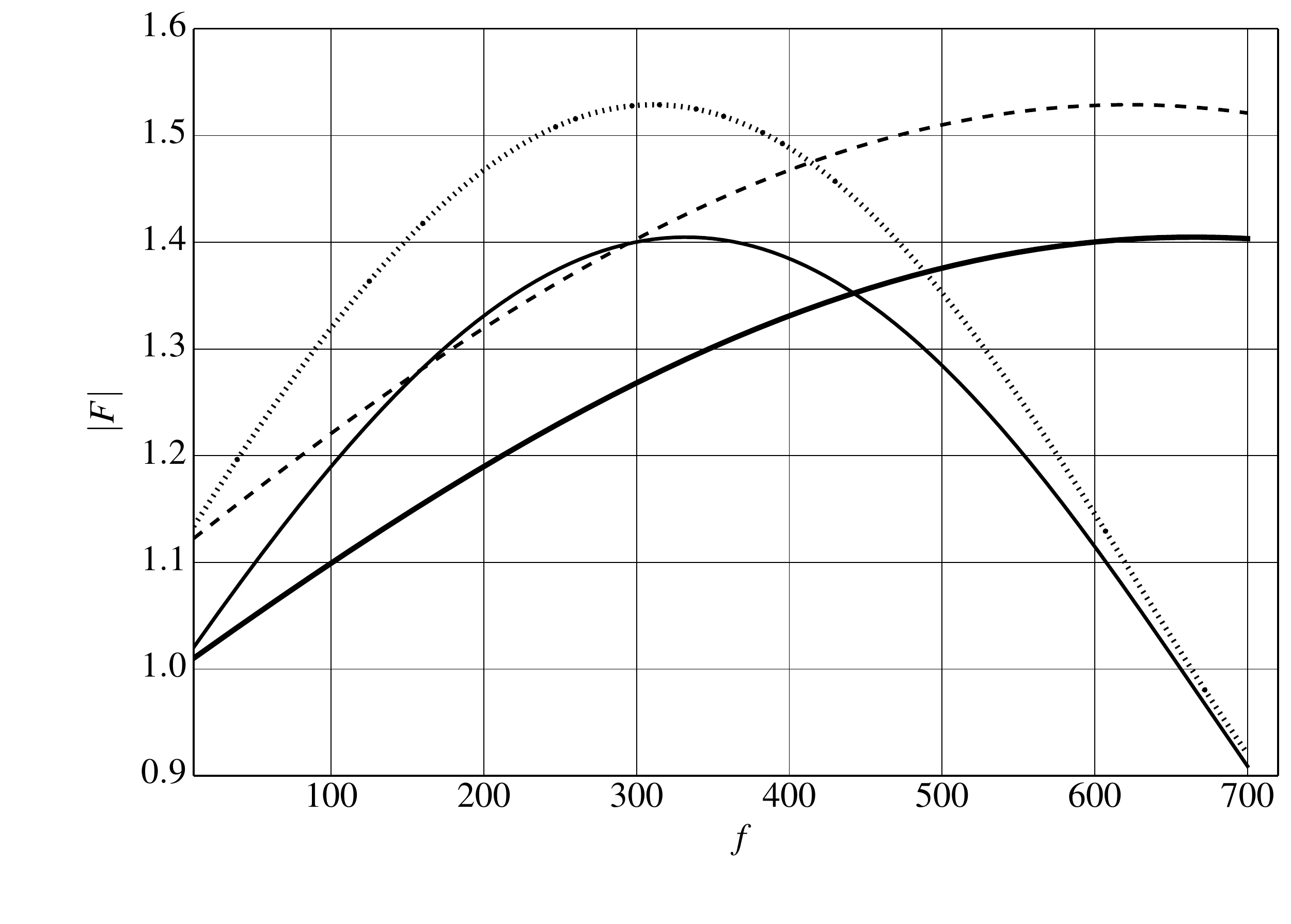}
 \includegraphics[width=8cm, height= 6cm]{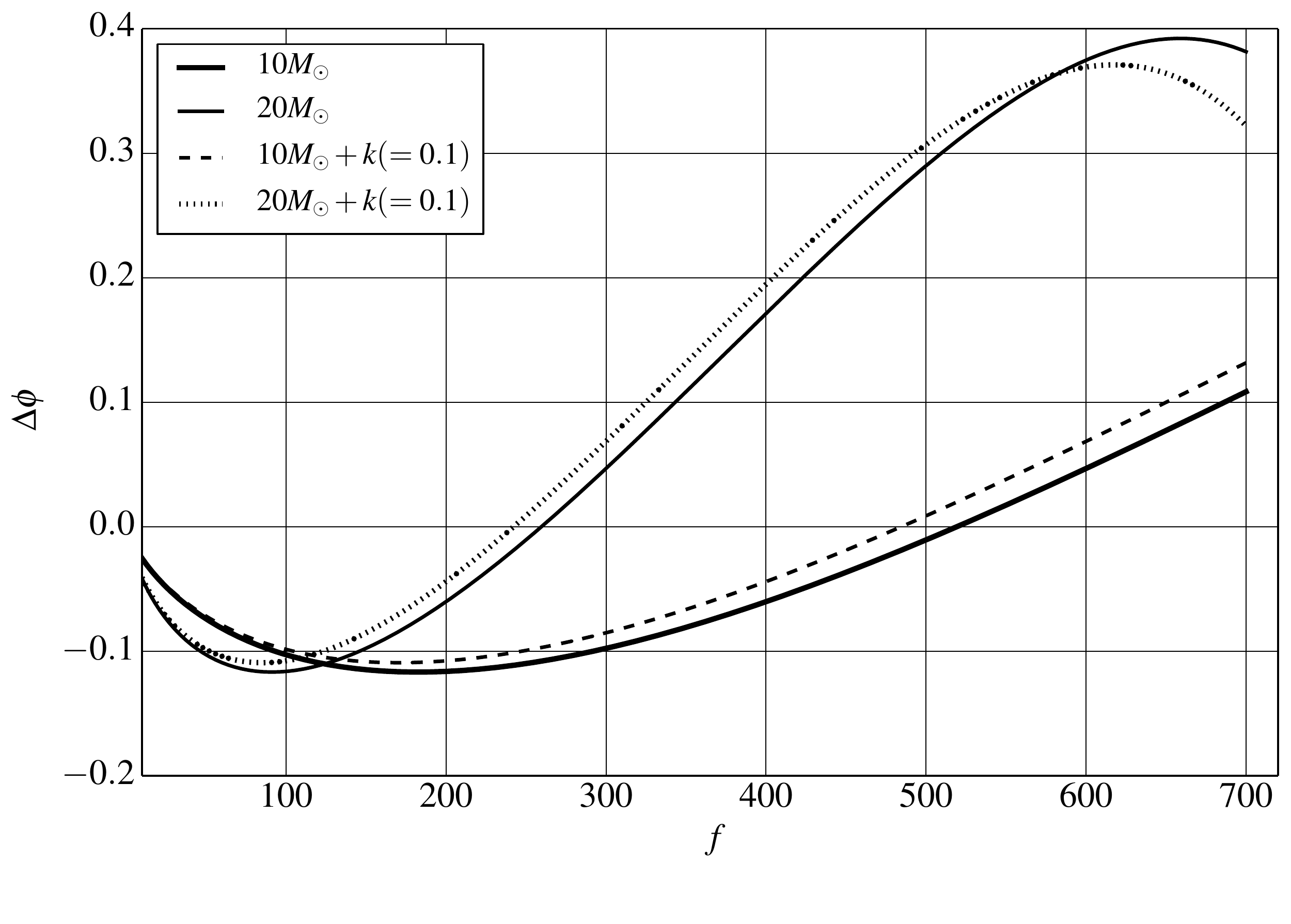} \\
 
 \includegraphics[width=8cm, height= 6cm]{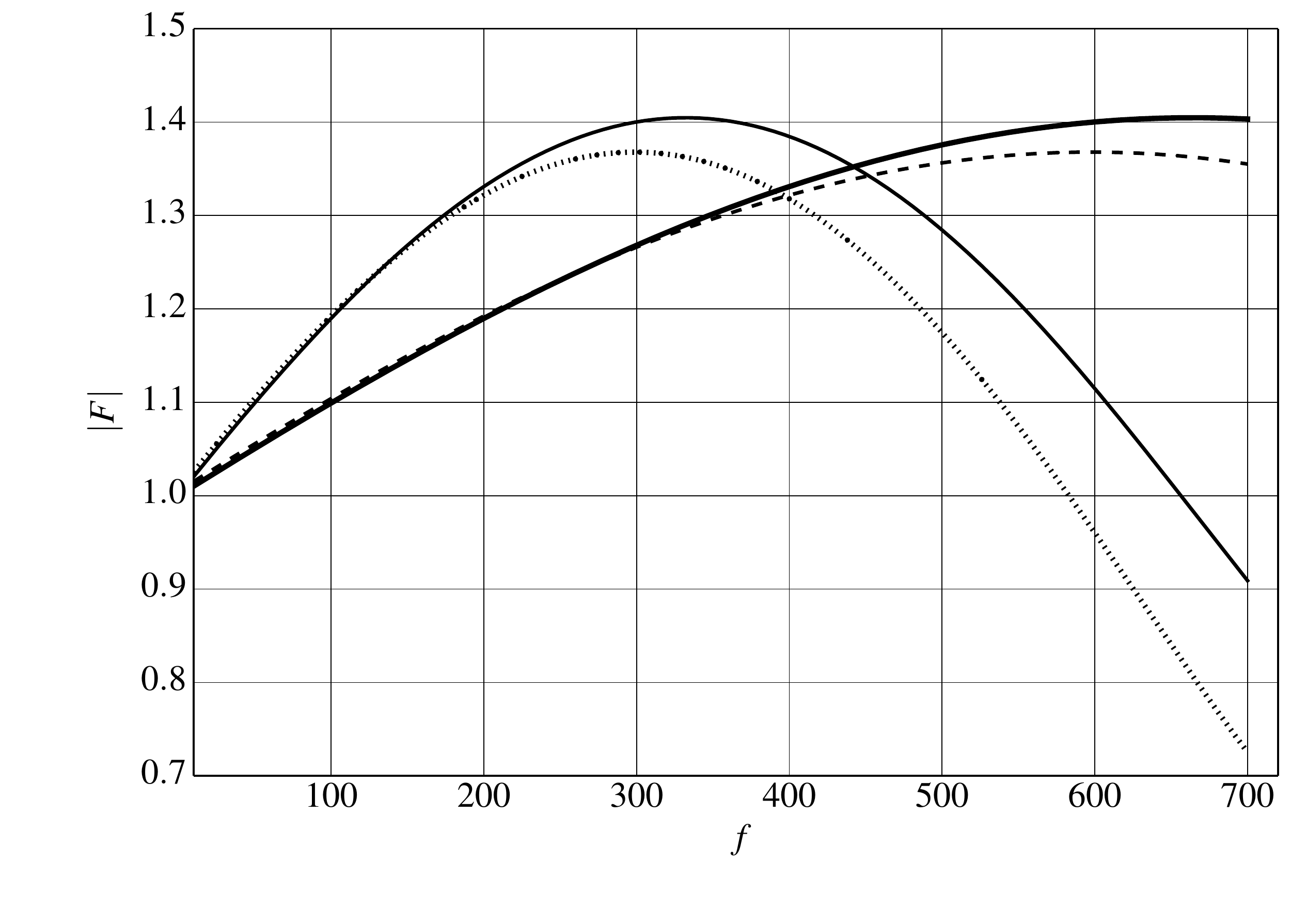}
 \includegraphics[width=8cm, height= 6cm]{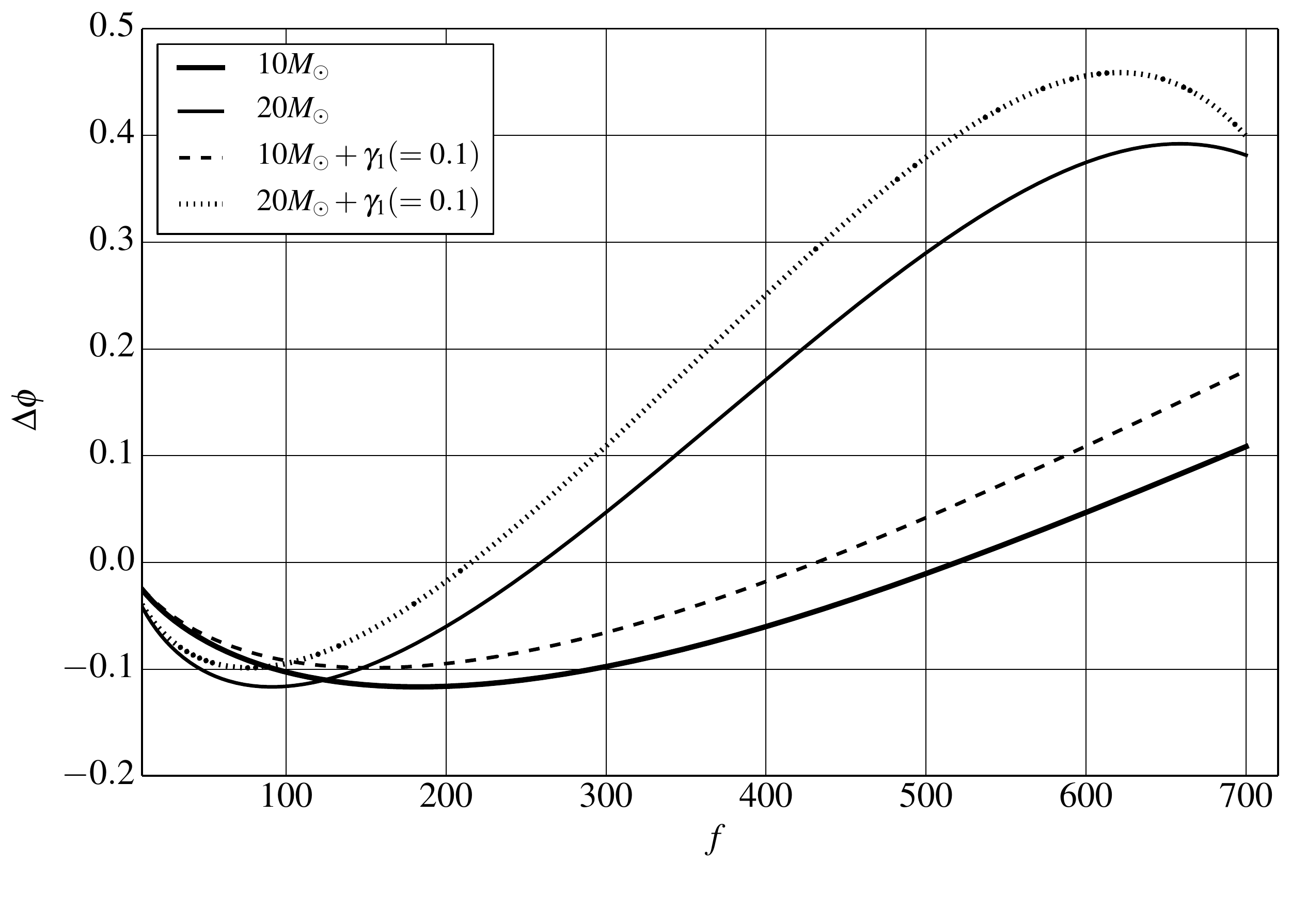} \\
 
 \includegraphics[width=8cm, height= 6cm]{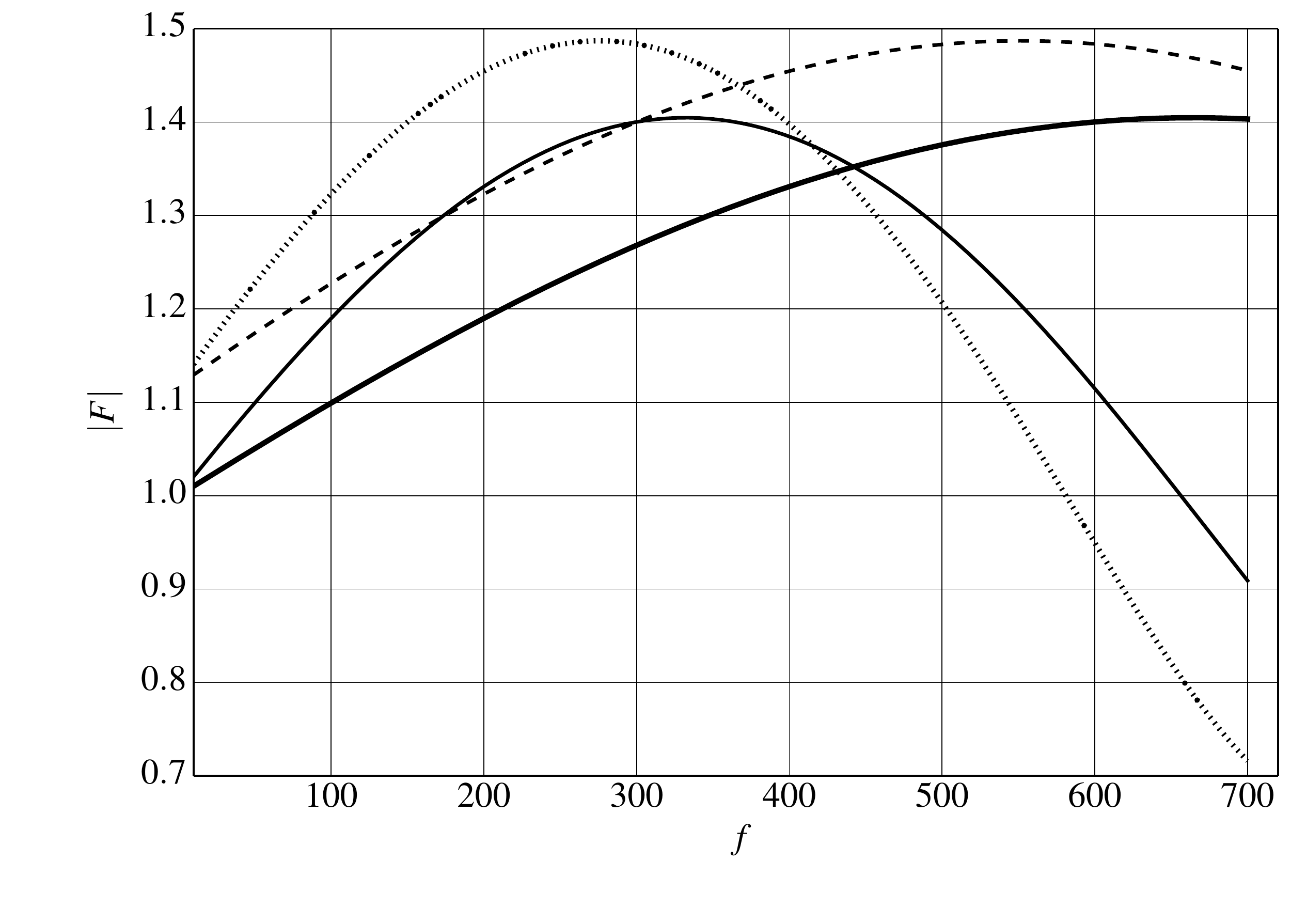}
 \includegraphics[width=8cm, height= 6cm]{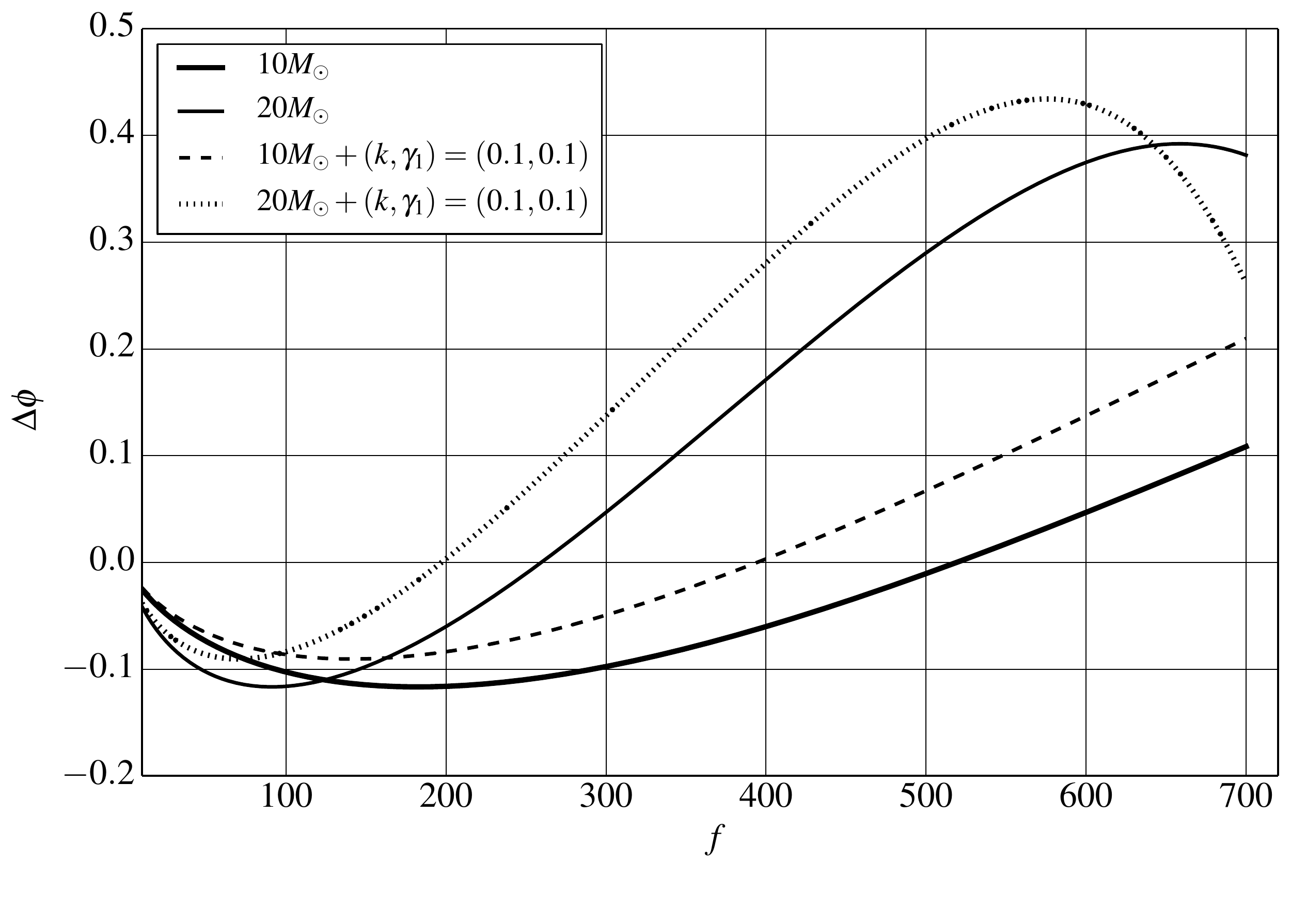}
 \caption{Micro-lensing within a strongly lensed image: The left column
   represents the amplification factor for two different micro-lenses
   within a strong lens represented by external shear and
   convergence. The right column shows the corresponding phase. For
   each row, the mass of the micro-lens and the values for external
   convergence and shear are mentioned in the right side of the
   respective row. The x-axis represents the frequency of
   gravitational waves in Hertz. The y-axis represents the
   amplification factor ($|F|$) and phase ($\triangle\phi$) in left
   and right panel, respectively. }  
 \label{fig:nkng}
\end{figure*}
%%%%%%%%%%%%%%%%%%%%%%%%%%%%%%%%%%%%%%%%%%%%%%%%%%%%%%%

In order to find out the amplification factor for different lens
models, we solve equation \eqref{eq:amplification_factor}
numerically.  
However, for a point mass lens, the solution of equation
\eqref{eq:amplification_factor} can be obtained analytically and is
given in terms of the hyper-geometric function, 
\begin{eqnarray}
F\left(\omega,y\right) &=& \exp\left[\frac{\pi
                           \omega}{4}+\frac{\textit{i}\omega}{2}
                           \Bigg\{\ln\left(\frac{\omega}{2}\right) -
                           2\phi_m\left(x_m\right)\Bigg\} 
                           \right]\nonumber\\  
&&
   \Gamma\left(1-\frac{\textit{i}\omega}{2}\right){}_{1}F_{1}
   \left(\frac{\textit{i}\omega}{2},1;\frac{\textit{i}\omega }{2}y^2
   \right),  
\label{eq:point_wave}
\end{eqnarray}
where $\omega=8\pi G (1+z_L)M_Lf/c^3$, $\phi_m\left(\mathbf{y}\right)
= \left(x_m-y\right)^2/2-\ln\:x_m$ with
$x_m=\left(y+\sqrt{y^2+4}\right)/2$.
Here we used Einstein radius as the scale length $\xi_0=\left(4GM_L D_L
  D_{LS}/c^2 D_S\right)^{1/2}$.
Using geometric optics approximation ($\omega \gg 1$), the above
equation reduces to, 
\begin{equation}
F\left(\omega,y\right)=\sqrt{|\mu_{+}|}-\textit{i}\sqrt{|\mu_{-}|}
\exp\left[{2\pi\textit{i}f\triangle t_d}\right] ,
\label{eq:point_geo}
\end{equation}
where $\mu_{\pm}$, are the amplification factor for primary and
secondary images for a point mass lens in geometric optics limit and
$\triangle t_d$ is the time delay between these two images. 
One can see the oscillatory behavior of the amplification factor at
higher frequencies.
To get the conventional expression for amplification factor, we need
to take a square followed by an average of equation
\eqref{eq:point_geo}.   
In our approach we have assumed a point source for gravitational
waves.
A further refinement where we take the finite size of the emitting
region can be made \citep{2006JCAP...01..023M}.

\section{Results}
\label{sec:results} 

In this section, we present the results of our study of gravitational
lensing of gravitational waves emitted from a coalescing binary.
For illustration we neglect the role of eccentricity of the orbit. 
This section is further divided into subsections, considering
different circumstances in which a gravitational wave can be affected
by a gravitational lens.

%%%%%%%%%%%%%%%%%%%%%%%%%%%%%%%%%%%%%%%%%%%%%%%%%%%%%%%
\begin{figure*}
  \includegraphics[width=8cm, height= 6cm]{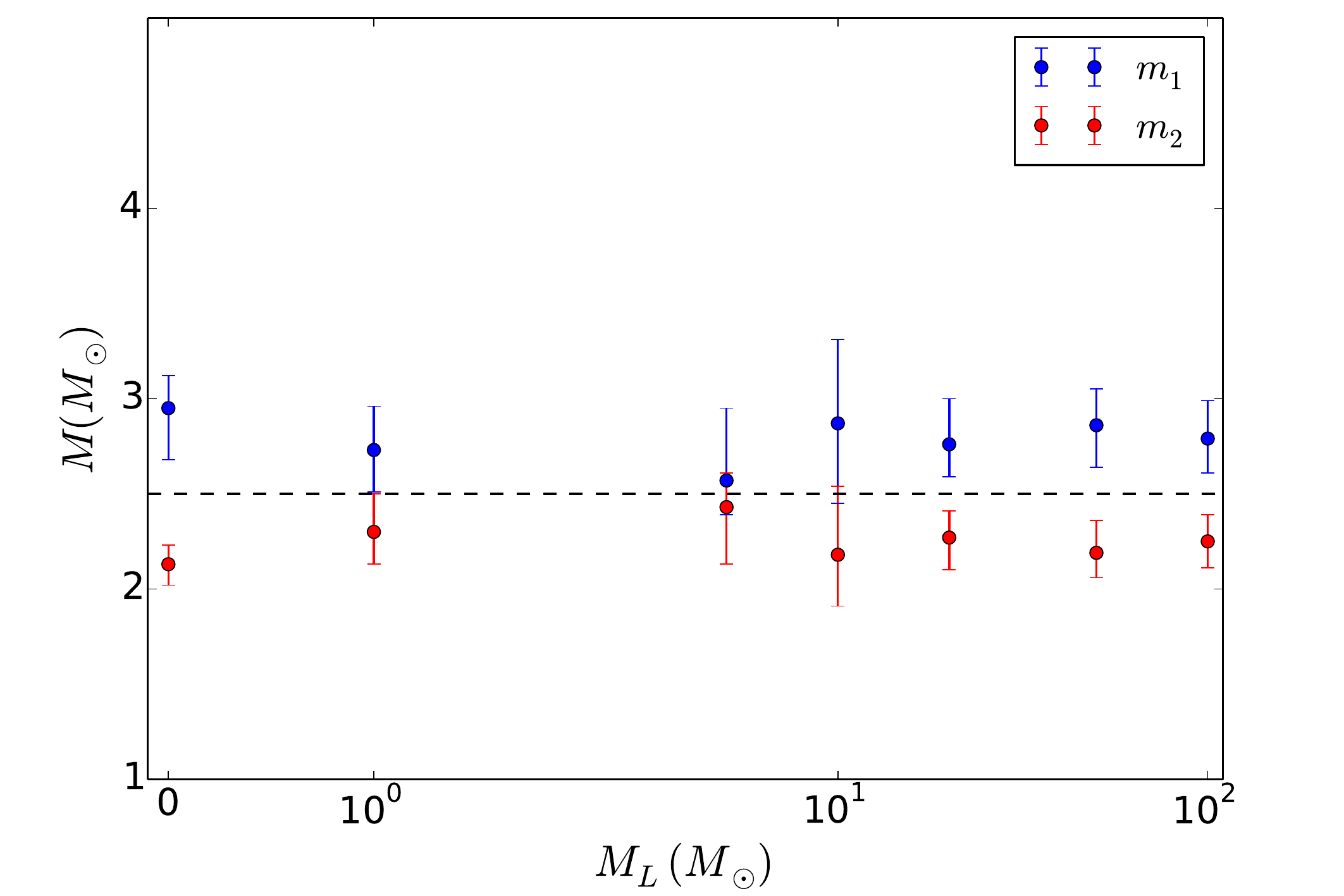} 
  \includegraphics[width=8cm, height= 6cm]{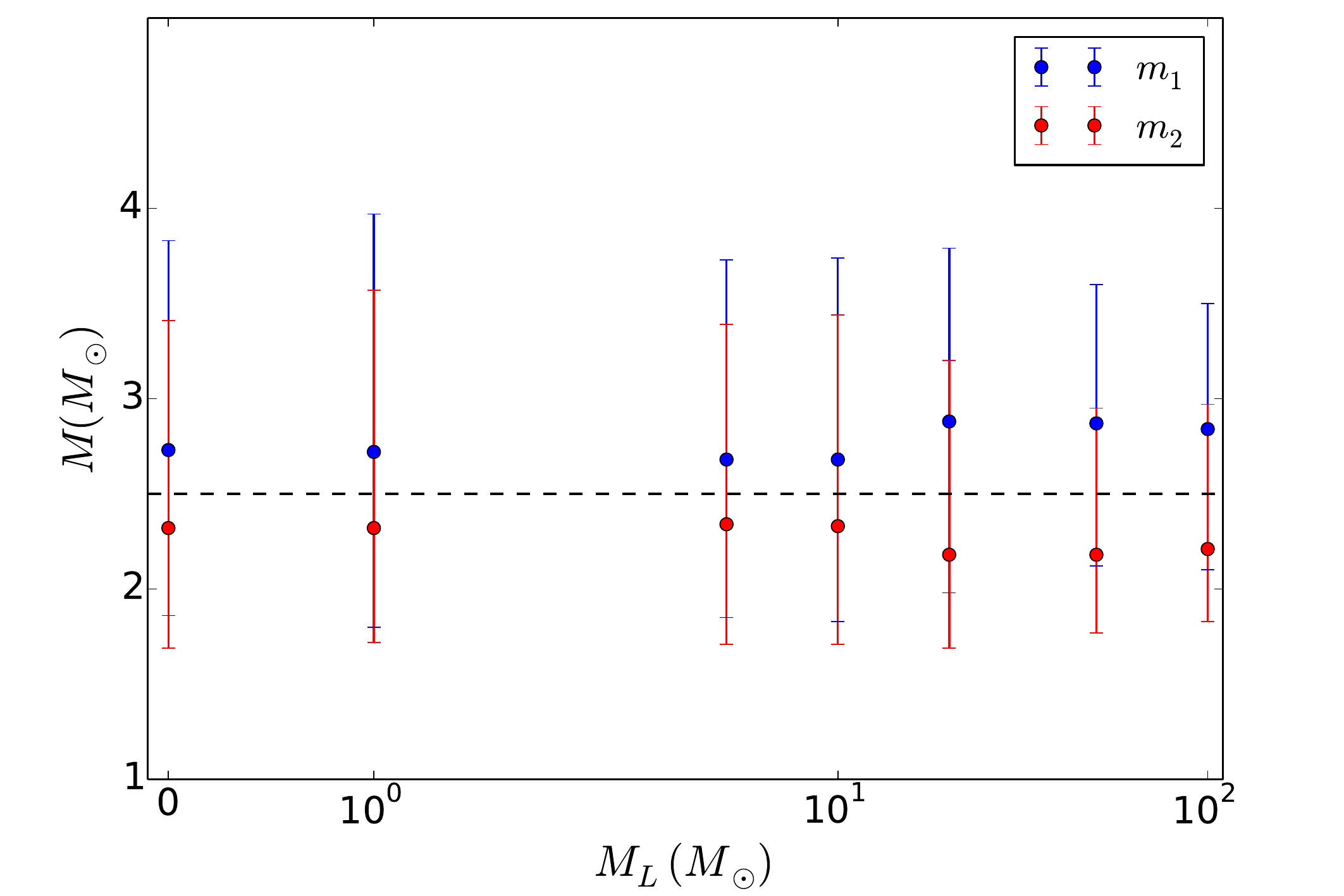} \\
  \includegraphics[width=8cm, height= 6cm]{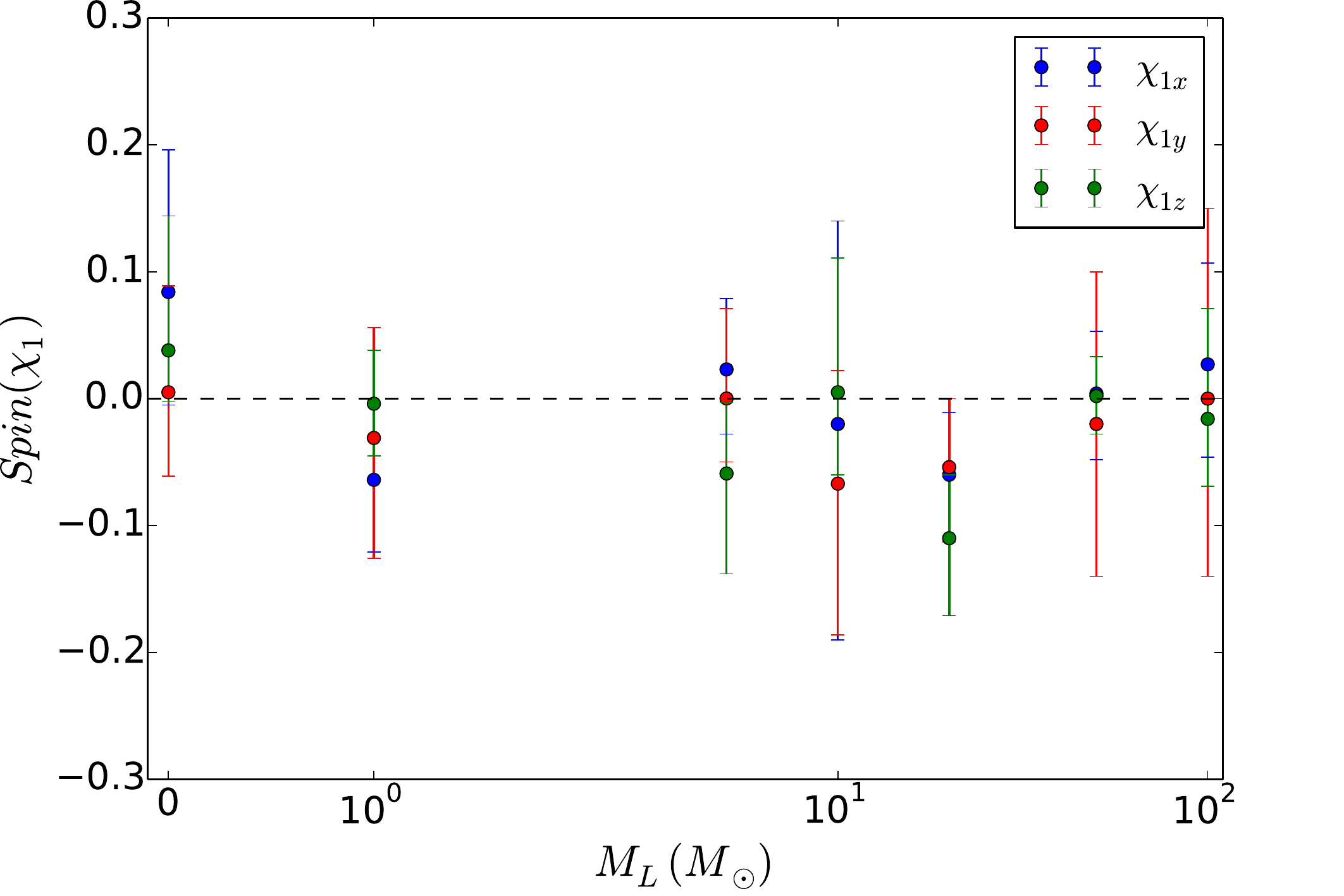} 
  \includegraphics[width=8cm, height= 6cm]{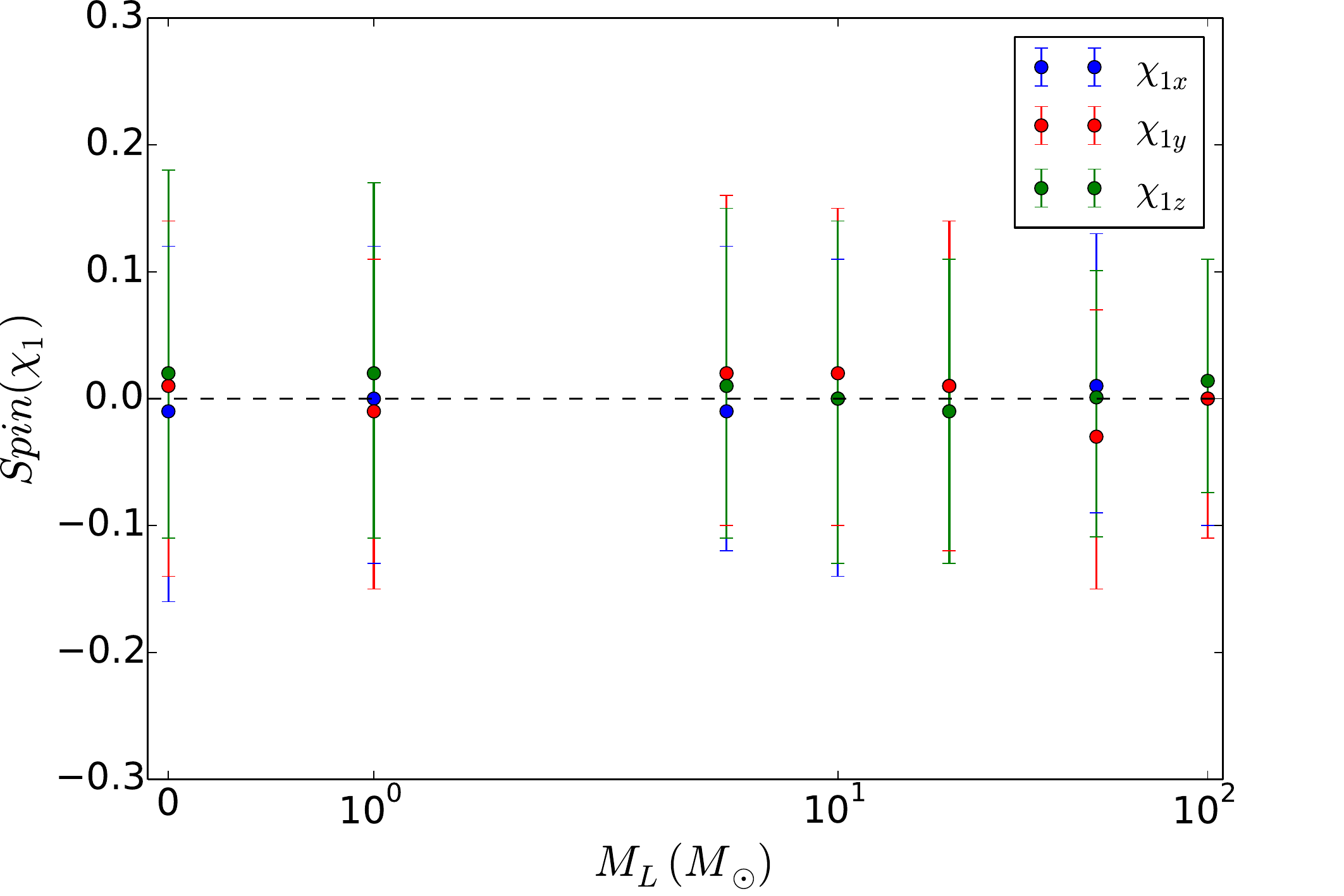}  \\
  \includegraphics[width=8cm, height= 6cm]{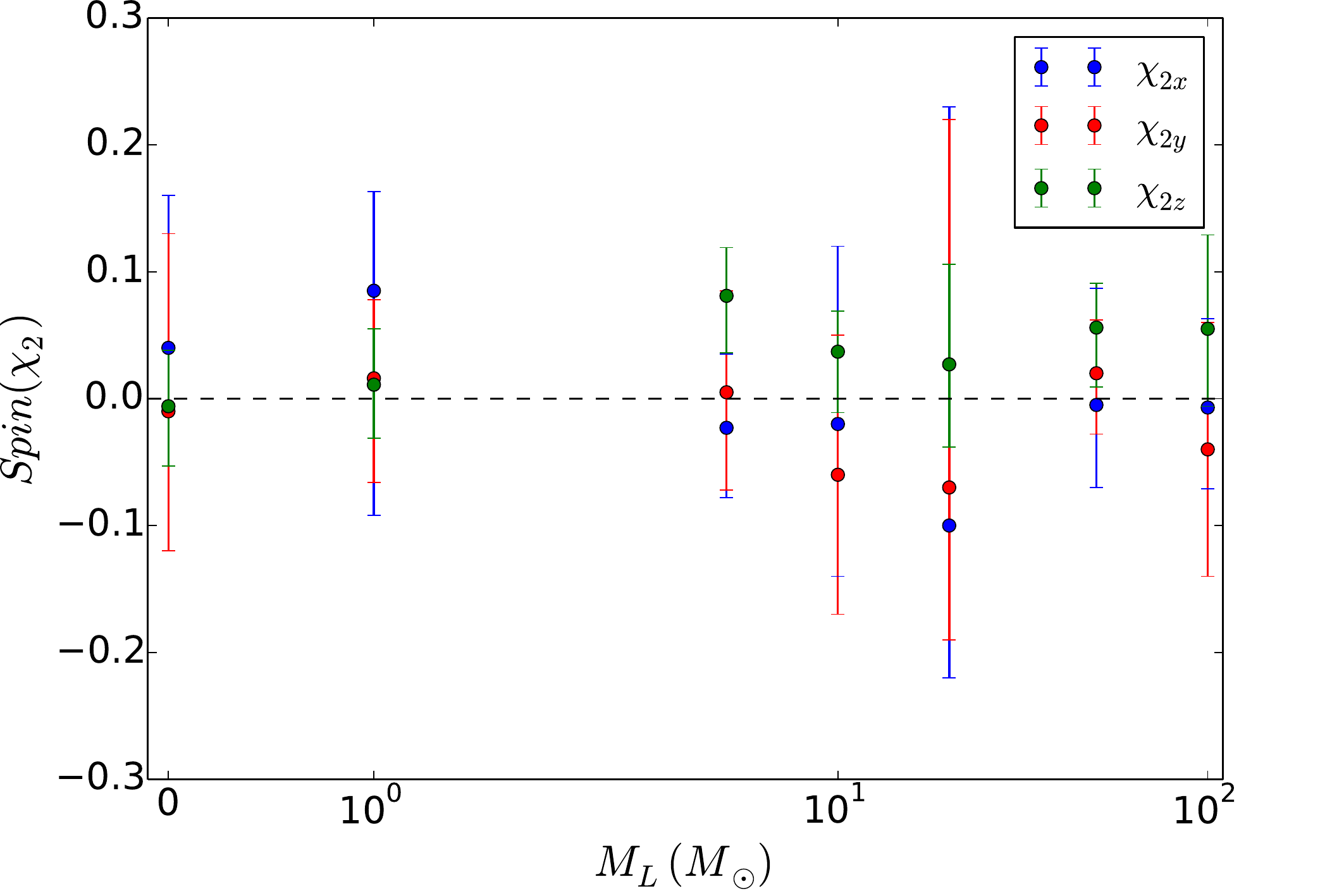} 
  \includegraphics[width=8cm, height= 6cm]{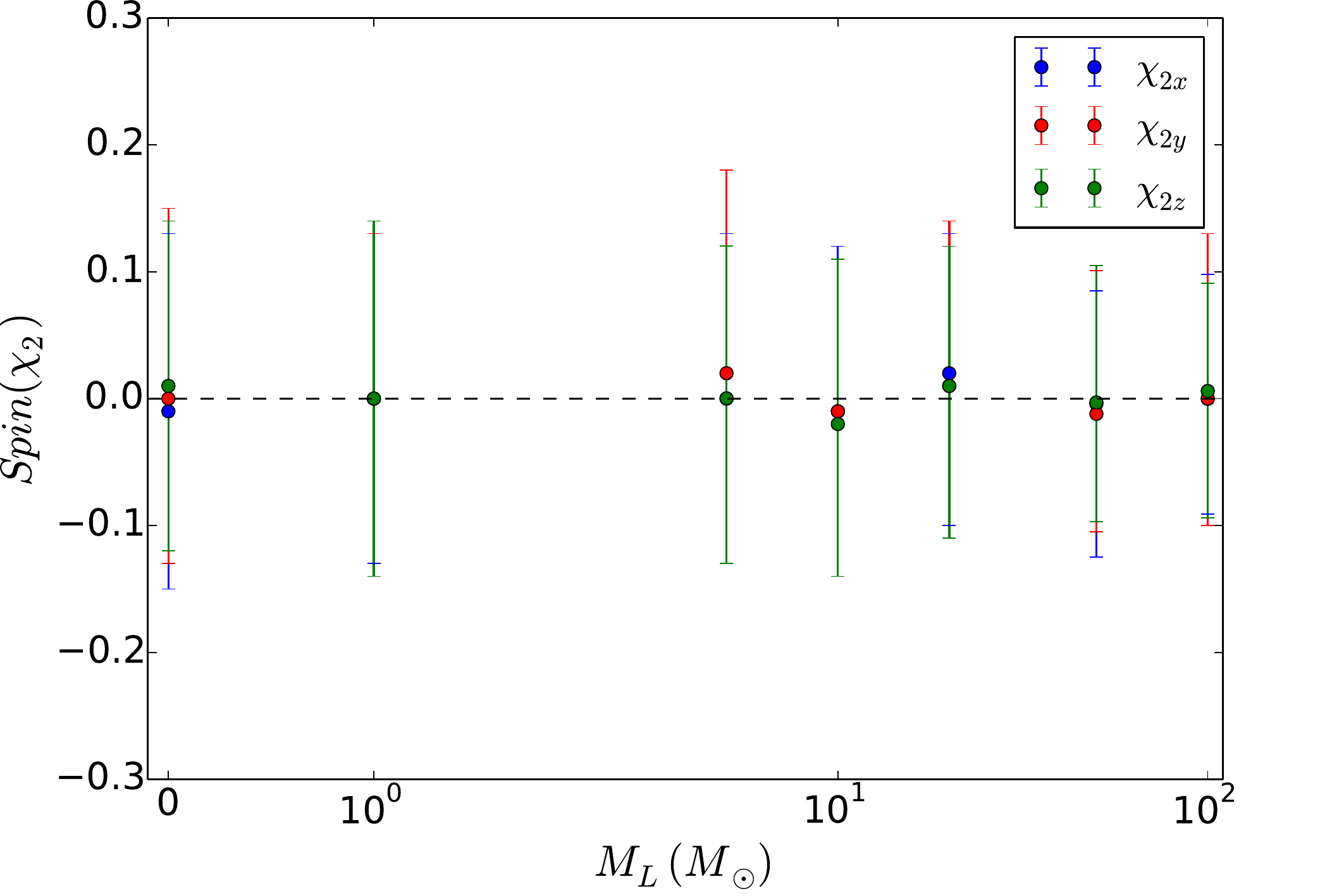} 
  \caption{Parameter estimation of a lensed GW signal at high and small SNR
  values: The left column shows the parameter estimation for a GW signal at 
  SNR value 34 (in unlensed case), whereas the right column shows the parameter
  estimation at SNR value 10 (in unlensed case). The x-axis represents the 
  microlens mass. The y-axis represents the different parameter values. In 
  different panels, the horizontal dotted line represents the input value of the
  corresponding parameters.
  The top row represents the mass estimation of binary components for different
  microlens values. The middle and bottom row represents the spin component
  estimation for first and second binary components, respectively. The error 
  bars show one sigma error in the estimation of different parameters.} 
 \label{fig:parameter_estimation}
\end{figure*}
%%%%%%%%%%%%%%%%%%%%%%%%%%%%%%%%%%%%%%%%%%%%%%%%%%%%%%%

%%%%%%%%%%%%%%%%%%%%%%%%%%%%%%%%%%%%%%%%%%%%%%%%%%%%%%%
\begin{figure*}
  \includegraphics[width=12cm, height= 6cm]{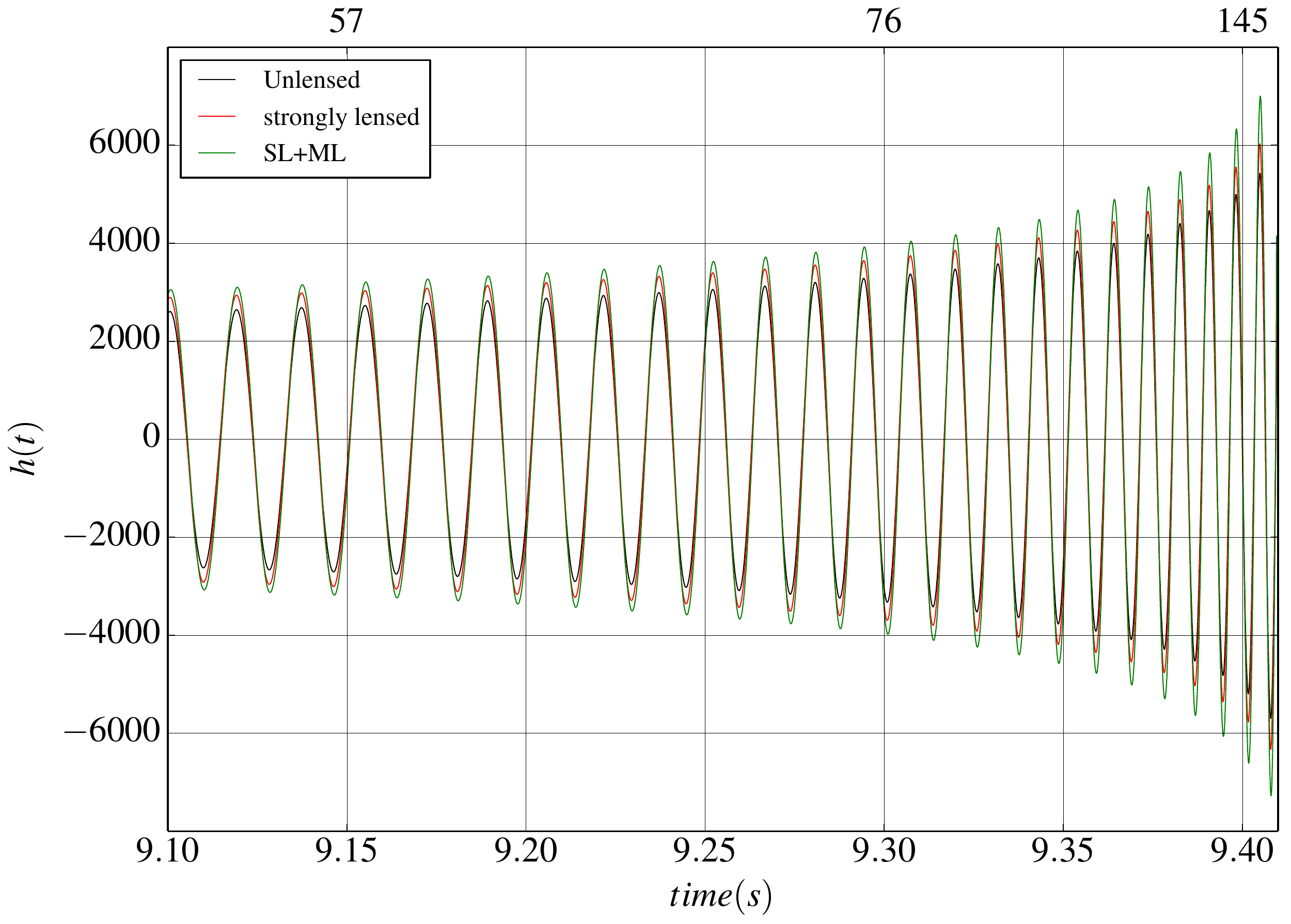} \\
  \includegraphics[width=12cm, height= 6cm]{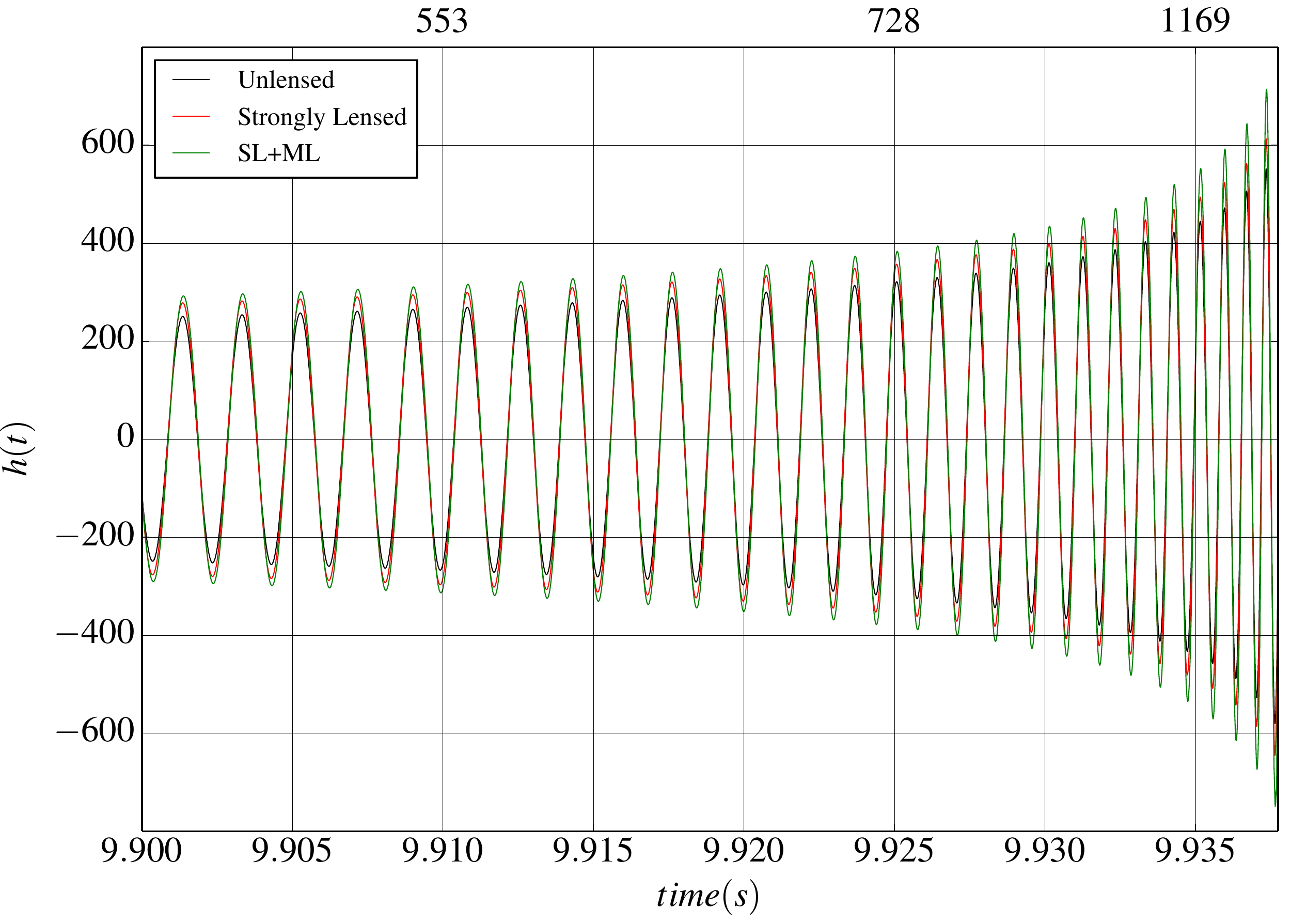} \\
  \includegraphics[width=12cm, height= 6cm]{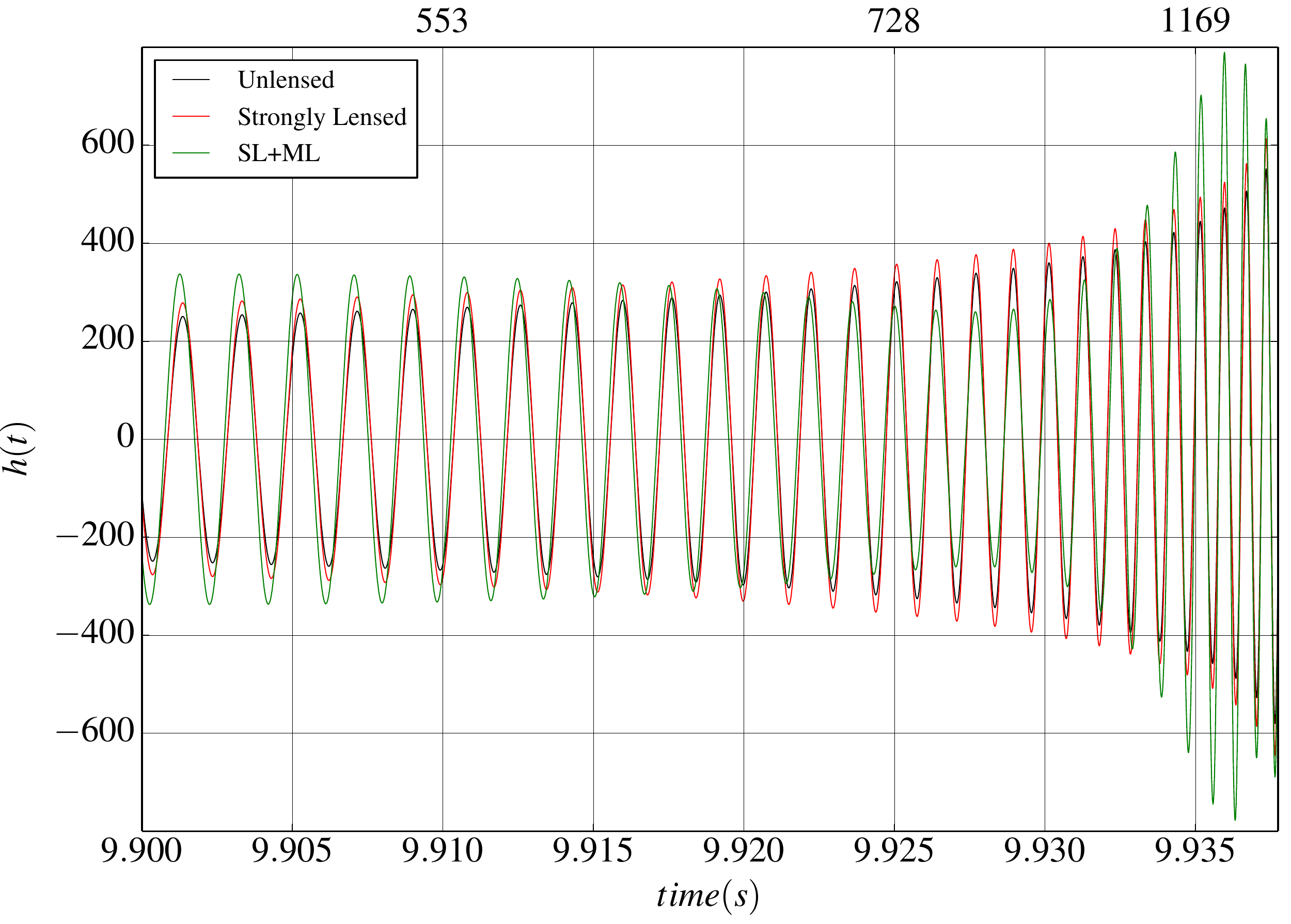} \\
  \caption{Effect of lensing on gravitational waveform in time domain.
    The top panel is for a $10$~M$_\odot + 10$~M$_\odot$ binary, the
    lower panels are for a $1$~M$_\odot + 1$~M$_\odot$ binary.
    The black curve shows unlensed signal, the red curve shows the
    effect of strong lensing (in the geometric limit), and the green
    curve shows the effect of micro-lensing in addition to strong
    lensing.  The top panel includes micro-lensing by a $10$~M$_\odot$
    lens, the middle panel includes micro-lensing by a $1$~M$_\odot$
    lens, and the lowest panel includes micro-lensing by a $20$~M$_\odot$
    lens. 
    In each panel the x$-$axis shows the time, the corresponding
    frequency of gravitational waves is marked on the top of each
    panel.
    Each panel shows the gravitational wave amplitude as a function of
    time (and frequency).
    The unlensed signal is amplified uniformly by strong lensing,
    independent of frequency. 
    However the inclusion of micro-lensing brings in the frequency
    dependence in the amplification where the amplitude of the signal
    gets amplified more at some frequencies.
    This can be seen in all the rows but it is seen most dramatically
    in the lowest row where this introduces a strong modulation.
    Such modulation is seen if we cover a wider range of frequencies,
    i.e., if the signal is from a lower mass source, or if the
    micro-lens is much more massive than a sun like star.
    There is also a frequency dependent phase shift that is obvious from
    lowest panel and can be read off in right panels of
    Figure~\ref{fig:nkng}.} 
 \label{fig:time_domain}
\end{figure*}
%%%%%%%%%%%%%%%%%%%%%%%%%%%%%%%%%%%%%%%%%%%%%%%%%%%%%%%

\subsection{Applicability of geometrical optics}

In case of lensing of electromagnetic radiation, due to the small
wavelength, use of geometric optics is adequate as corrections from
the wave optics are negligible \citep{Schneider_1992}.  
However, the wavelength of gravitational waves that can be detected
with existing and future detectors is much larger and is comparable
to the relevant scales of some gravitational lenses.
Therefore in order to discuss the lensing of gravitational waves, we
need to ascertain whether wave effects are significant or not?
We find that for wavelengths equal to or larger than the path
difference between the multiple images of the source, one should use
wave optics instead of geometric optics.  
Taking equation \eqref{eq:amplification_factor} into account, one can
write the above condition in mathematical form
as~\citep{Takahashi_2017},  
\begin{equation}
2\pi f t_d \lesssim 1,
\label{eq:condition}
\end{equation}
where $f$ is the frequency of the gravitational waves and $t_d$ is the
time delay between the images. 
Figure \eqref{fig:cutoff_frequency} shows the plot of the equation
\eqref{eq:condition} for a point mass lens.  
The thin solid line denotes the condition $2\pi f t_d = 1$. 
Below this line one should use the wave optics results as the
wavelength of the gravitational waves is much larger than the path
difference between the two images, which is of the order of the
Schwarzschild radius of the lens.  
Above but near this line, the correction terms from diffraction
effects continue to be non-negligible.  
As one moves away from this line, the correction terms due to
diffraction effects become negligible, and finally one can use
geometric optics if the wavelength of the gravitational wave is much
smaller than the Schwarzschild radius of the lens.
Geometric optics is a good approximation if we are a factor of at
least a few above this line.

The light and dark shaded regions represent the frequency ranges
covered by LISA and LIGO, respectively. 
The horizontal lines (dotted, dashed, thick solid line) represents the
frequency of the gravitational waves at the innermost stable circular
orbit (ISCO) emitted by the source binaries of masses, $100
M_\odot+100 M_\odot$, $10M_\odot+10M_\odot$, $1M_\odot+1M_\odot$,
respectively.  
For the purpose of illustration we have neglected the effects of
eccentricity and post-Newtonian corrections while computing the
relevant frequency. 
The corresponding time delay is shown in the right side of the plot. 
One can see that for most of the relevant sources, the frequency
of gravitational waves at ISCO lies in the LIGO band and the wave
effects are only important in case of a micro-lensing scenario
\citep{2018PhRvD..98j3022C}.  

The frequency dependent amplification and phase shift leads to
chromatic effects and this is the key signature of wave effects.
These effects modify the waveform from coalescing binaries as the
frequency keeps on increasing with time in the signal.
The changes can be large enough to lead to mis-identification of
source properties.
\citet{2018PhRvD..98j3022C} have shown that disentangling the effects
of micro-lensing is feasible for detections with a high signal to noise
ratio ($SNR \geq 30$).  

\subsection{Micro-lensing Effects}
\label{sub:micro}

The possibility of strong lensing of gravitational waves gives us the
opportunity to observe the same source more than once due to the
nonzero time
delays
\citep{2013JCAP...10..022P,smith_2018,li_2018,Haris_2018,Hannuksela_2019}.  
In general, these different signal have different constant
amplification factor, depending on the geometry of the lensing system.  
However, as we know smaller mass compact objects (stars, stellar
remnants, black holes) are a part of galaxies and can further affect
the gravitational wave signal via micro-lensing.  
The lensing due to these micro-lenses can also help us in observing
intermediate mass black holes (IMBH) \citep{lai_2018}. 
As mentioned above, in LIGO frequency range, only micro-lensing
($M_{lens} \leq 10^3$~M$_\odot$) can give rise to the significant
frequency dependent effects. 

For simplicity we only consider a single object as a potential
micro-lens. 
In order to calculate the effect of micro-lensing, we choose the object 
to be located at the origin. 
The effect of the galaxy as a whole does not vary significantly over
the relevant length scale of a star.  
As a result, the effect of the galaxy can be describe in terms of
constant external convergence ($\kappa'$) and constant shear
($\gamma_1'$, $\gamma_2'$).  
Due to the presence of external effects, the resultant lensing
potential of the star will be modified and can be written
as~\citep{Schneider_1992}:
\begin{eqnarray}
\psi\left(x_1,x_2\right) &=& \ln\left(\sqrt{x_1^2 + x_2^2}\right) +
                             \frac{\kappa'}{2}\left(x_1^2 +
                             x_2^2\right)\nonumber\\ 
&& + \frac{\gamma_1'}{2} \left(x_1^2 - x_2^2\right) + \gamma_2' \: x_1
   x_2.
   \label{eq:external_effects}
\end{eqnarray}
Using equation \eqref{eq:amplification_factor} and
\eqref{eq:external_effects}, we can calculate the effect of the star
on the amplification factor as well as the phase of the gravitational
wave signal.  

Here we show results for three different possibilities: non-zero
convergence, non-zero shear, and non-zero convergence and shear for a
lens at redshift $z=0.05$ and the source positioned at
$\left(y_1,y_2\right) = \left(1,0\right)$ at redshift $z=0.2$.  
The mass of the micro-lens is fixed to $10M_\odot$ and $20M_\odot$. 
As one can see from figure (\ref{fig:nkng}), the non-zero value of
convergence (shear) introduces a frequency dependence in the
amplification factor as well as in the phase (these would have been
independent of frequency in the absence of micro-lensing). 
Further, this frequency dependence is highly sensitive to the lens
parameters ($M,\kappa,\gamma_1,\gamma_2$).   
For example, if we change the mass on the micro-lens from $10M_\odot$
to $20M_\odot$, the amplification and phase pattern show significant
changes.  
Similarly, if we consider a different combination of
$\left(\kappa,\gamma_1,\gamma_2\right)$ which is equivalent to
shifting the position of the micro-lens, the amplification pattern
changes (see figure \ref{fig:nkng}).
The effect on the chirp signal from coalescing binaries is to modulate
the amplitude as we have a frequency dependent amplification, and also
to introduce time varying phase shifts.

The effect of these oscillations on parameter estimation is 
  shown in figure~(\ref{fig:parameter_estimation}).
  We have considered the effect of the microlens and ignored the
  effect of the strong lens galaxy as that does not introduce any
  frequency dependent variations in the signal. 
  The left and the right column represents the parameter estimation of
  different binary parameters for a $2.5$~M$_\odot + 2.5$~M$_\odot$
  binary for different SNR values.
  The SNR value in the left column is 34 for the unlensed signal,
  whereas the SNR value in the right column is 10 for the unlensed
  signal.
  These cover a reasonable range in SNR and give us a glimpse of what
  to expect. 
  Here we used PyCBC for parameter
  estimation~\citep{2019PASP..131b4503B}.
  As one can see in the right column of
  figure~(\ref{fig:parameter_estimation}), at small SNR values
  ($\sim$10), the estimated values of various binary parameters do not
  differ significantly within one sigma errors.
  However, if the SNR value is larger than $30$, then the microlensing
  effects on the parameter estimation can be distinguished within one
  sigma errors.
  Hence, the effect of microlensing is to modify the estimated values
  of binary parameters.
  This can be seen in the middle and bottom panel of the left column
  of figure~(\ref{fig:parameter_estimation}), where a non-zero spin is
  estimated even though the input spin is zero.
  It is interesting to note that the extent of the allowed region
  remains similar to the case of no micro-lensing (marked by $0$ on
  the x-axis) but the allowed region moves around with variation in
  the microlens mass. 
  These modifications can be more significant if we also take into
  account the impact of the host galaxy as the effect of the host
  galaxy can further increase the SNR value(figure~\ref{fig:nkng}).

This introduces a severe problem in identifying different counterparts
of a strongly lensed GW signal as the micro-lensing can introduce a
different type of frequency dependence in different images.   

Keeping source properties the same, if we explore the effects with
changes in the mass of the micro-lens, we find that the form of
variation of amplification and phase shift with frequency is 
monotonic in the LIGO range for micro-lenses with a smaller mass
($\sim $M$_\odot$), whereas it undergoes oscillations for a larger
mass ($> 10^2$M$_\odot$).  
This is to be expected as with larger masses we gradually approach the
geometric limit.

For better visualization, figure~(\ref{fig:time_domain}) represents
the lensed waveform in the time domain.
We show the expected signal in the final moments before coalescence
for three cases.
We see here that for the low mass binary we can probe gravitational
waves of a higher frequency and hence we are able to see the effect of
oscillation in amplification.
No such feature is seen in the signal from the higher mass binary as
it does not reach higher frequencies.
However, a frequency dependent amplification is seen in all cases.
There is a frequency dependent phase shift that is not obvious from
these figures but which is large enough to mislead in the mapping from
the signal to the source parameters.
As micro-lensing does not affect each sight line in
the same manner, a comparison of the red and the green color waveform
in bottom panel of figure~(\ref{fig:time_domain}) is a representation
of the differences that we may encounter due to this effect.
It is clear from here that it may be difficult to identify signal from
different strongly lensed images as coming from the same source due to
the variations introduced by micro-lensing.

The probability of a micro-lensing event depends on many factors but it
is not small.
We are doing a detailed estimation but a preliminary analysis suggests
that a few percent of the images are likely to be affected by
micro-lensing. 
This order of magnitude analysis is done for an elliptical
  galaxy acting as a gravitational lens lens with de Vaucouleurs
  profile along with the Salpeter initial mass function.  
  The average microlensing event rate for such a galaxy lens is of the
  order of $10^{-2}$.
  If we specifically consider the microlensing near half-light radius,
  then the value of optical depth is larger than the average value:
  this is perhaps more relevant as the impact parameter in most cases
  is of this order or smaller.  
  Similarly, if we consider microlensing due to an edge-on spiral
  galaxy, then the probability of microlensing can further rise as the
  projected density of the microlenses within the disk is much
  higher.
  On the other hand, the probability for microlensing for off plane
  images is much lower in this case.

\section{Conclusions}
\label{sec:conclusions}

We have described the effects of micro-lensing  alone and in case of a 
strongly lensed GW signal.  
As shown above, micro-lensing is the only way in which lensed signal in
LIGO frequency band can get frequency dependent effects.
The micro-lens may reside in our galaxy or it may be embedded in a
bigger lens, e.g., an intervening galaxy.
The effects of micro-lensing depend strongly on the lens parameters. 
For large values of SNR ($>30$), as we change any of 
the lens parameters, the variation of
amplification and phase with frequency changes significantly.  
Because of the possibility that different counterparts of strongly
lensed GW signal can get affected differently by micro-lensing, or only
one image may be affected significantly, it introduces a real
challenge in identifying the counterparts of the strongly lensed GW
signal. The other difficulty in identifying the 
different counterparts of a strongly lensed GW signal is the poor 
localization of the GW signal. However, this identification can be
possible if SNR is large enough (more than about 30) but most 
detections fall below this threshold. Apart from that, if the 
magnification factor due to the strong lens is large, then the 
possibility of microlensing 
is unavoidable. In such a case, even a few solar mass microlens can 
introduce significant distortions 
in the observed GW signal~\citep{Diego_2019}.

The challenge can be turned into an opportunity if a counterpart is
observed in the electromagnetic radiation as this may permit better
localization and identification of the images.
In such cases we can recover information about the wave effects and
parameters of the micro-lens.
Here, the sensitivity of the amplification and phase difference to
parameters of the lens are useful features.
In addition, time delay measurements with optical counterparts can be
used to infer a lot of information about the lens and cosmological
parameters \citep{2013JCAP...10..022P,2018arXiv180906511H}.

Gravitational lensing in the LISA band by much heavier lenses retains
effects of the wave nature and hence it should be possible to detect
these \citep{2019arXiv190306612L}.
However, lens masses required are in a range where few potential lens
candidates are available.
We propose that super massive black holes in intervening galaxies may
produce effects that can be tracked in the LISA band.
Observations here are sensitive to the mergers of very massive black
holes, or to other coalescing binaries at epochs much before the final
merger.
In the latter case, superposition of signals with different time
delays can produce modulation
effects~\citep{Takahashi_2003,Sereno_2010}.

Micro-lensing can affect gravitational waves from distant sources even
in the absence of strong lensing.
This may happen due to a lens in an intervening galaxy or a lens in
the Galaxy.
The probability for this to happen is small in most directions but is 
relatively \citep{2005MNRAS.362..945W} very high in
the galactic plane region.  
Given that this can affect signal from sources in more than $10\%$ of the
sky with a high probability, effects of micro-lensing need to be taken
into account seriously.
It has been shown \citep{2018PhRvD..98j3022C} that for detections with
a high SNR it is possible to disentangle the micro-lensing effects and
source parameter determination.
However, it is important that for other events we budget for the
uncertainty that is introduced because of the likelihood of
micro-lensing.

\section*{Acknowledgements}

AKM would like to thank CSIR for financial support through research
fellowship No.$524007$.
We thank the anonymous referee for useful comments. 
We acknowledge the use of IISER Mohali HPC facility.
This research has made use of NASA's Astrophysics Data
System Bibliographic Services.

\bibliographystyle{mn2e}

%\bsp
\label{lastpage}
\end{document}